\documentclass[preprint,showpacs,preprintnumbers,eqsecnum,floatfix]{revtex4}

\usepackage{graphicx}% Include figure files
\usepackage{bm}% bold math

\def\lesssim{\ \raise.3ex\hbox{$<$}\kern-0.8em\lower.7ex\hbox{$\sim$}\ }
\def\gesim{\ \raise.3ex\hbox{$>$}\kern-0.8em\lower.7ex\hbox{$\sim$}\ }

\begin{document}
%\preprint{}

%Title of paper
\title{Tunneling properties of a bound pair of Fermi atoms in an optical lattice}
\author{Y. Ohashi$^{1,2}$}
\affiliation{
$^1$Faculty of Science and Technology, Keio University, Hiyoshi, Yokohama, 223-8522, Japan\\
$^2$CREST(JST), 4-1-8 Honcho, Saitama 332-0012, Japan}
\date{\today}

\begin{abstract}
We investigate tunneling properties of a bound pair of Fermi atoms in an optical lattice, comparing with results obtained in an attractive Hubbard model. In the strong coupling regime of the Hubbard model, it has been predicted that the motion of a bound pair between lattice sites is accompanied by virtual dissociation. To explore the possibility of this interesting phenomenon in optical lattice, we calculate molecular wavefunction in a cosine-shape periodic potential. We show that the molecular tunneling accompanied by dissociation occurs in the intermediate coupling regime of the optical lattice system. In the strong coupling regime, in contrast to the prediction in the Hubbard model, the bound pair is shown to tunnel through lattice potential {\it without} dissociation. As a result, the magnitude of molecular band mass $M$ remains finite even in the strong coupling limit, which is in contrast to the diverging molecular mass in the case of the Hubbard model. Including this finite value of molecular band mass, we evaluate the superfluid phase transition temperature $T_{\rm c}$ in the BEC limit of the optical lattice system, where the Hubbard model gives $T_{\rm c}=0$ due to the diverging molecular mass. 
\end{abstract}


\maketitle

%%%%%%%%%%%%%%%%%%%%%%%%%%%%%%%%%%%%%%%%%%%%%%%%%%%%%%%%%%%%%%%%%%%%%%%%%%%%%%%
\section{introduction}
In 2006, the superfluid state was realized in a $^6$Li Fermi gas loaded on an optical lattice\cite{MIT1,MIT2}. In the optical lattice, atoms feel a periodic potential produced by standing wave of laser light\cite{Pita}. Thus, together with a tunable pairing interaction associated with a Feshbach resonance\cite{Tim,Hol,Ohashi}, we can now study lattice effects on Fermi superfluids in the BCS-BEC crossover region. Since Fermi gases in optical lattices are similar to conduction electrons in metals, superfluid lattice Fermi gases may be also useful for the study of metallic superconductivity. 
\par
Various effects of optical lattice on superfluid Fermi gases have been studied theoretically\cite{Zoller,Orso1,Orso2,Ho,Machida,Tamaki,Kawakami,Wakanabe,Kok}. Among them, one important effect is an anisotropic Fermi surface. When the lattice potential is strong, the lattice Fermi gas is expected to be close to the Hubbard model, consisting of nearest-neighbor hopping $-t$ and on-site pairing interaction $-U$. In this case, the Fermi surface in the cubic lattice has the nesting property at the half-filling, characterized by the nesting vector ${\bf Q}=(\pi/d,\pi/d,\pi/d$), where $d$ is the lattice constant. This perfect nesting induces strong density wave fluctuations, the strength of which is comparable to pairing fluctuations\cite{Micnus,Tamaki}. As a result, the coexistence of superfluid state and density wave state is realized in the half-filling case\cite{Micnus,Tamaki,Machida,Kawakami}. Since the competition between these two kinds of fluctuations is absent in a uniform Fermi superfluid, this coexistence phenomenon is characteristic of lattice Fermi superfluids. We note that the anisotropic Fermi surface has been recently observed in a $^{40}$K lattice Fermi gas\cite{Kohl}.
\par
Besides this, kinetic properties of bound pairs (Cooper pairs) are also strongly affected by optical lattice potential. In the strong coupling regime of the Hubbard model, it has been shown that the hopping of a bound pair between lattice sites is accompanied by virtual dissociation\cite{Nozieres}. This comes from the fact that the ordinary Hubbard model consists of the {\it atomic} hopping and on-site interaction. Namely, when the bound pair moves between lattice sites, each atom in this molecule has to move one by one. This tunneling process naturally leads to the enhancement of molecular band mass in the strong coupling regime as $M\propto E_{\rm bind}$\cite{Nozieres}, where $E_{\rm bind}$ is the molecular binding energy. Since $E_{\rm bind}$ diverges in the strong coupling limit, the molecular mass $M$ also diverges, leading to the vanishing superfluid phase transition temperature $T_{\rm c}$\cite{Micnus,Tamaki,Levin2}.  We note that the molecular mass equals twice the atomic mass in the strong coupling regime of a {\it uniform} Fermi superfluid with no optical lattice, leading to the finite value of $T_{\rm c}=0.218T_{\rm F}$\cite{Ohashi,Nozieres,Randeria} (where $T_{\rm F}$ is the Fermi temperature). 
\par
The Hubbard model is usually expected to be valid for the optical lattice system when the lattice potential is strong. However, Orso and co-workers\cite{Orso1,Orso2} recently studied a bound state problem in a realistic cosine-shape optical lattice potential, and showed that the molecular band mass $M$ actually does not diverge but remains finite even in the strong coupling limit. Their result indicates that the Hubbard model is {\it not} valid at least for the strong coupling limit of optical lattice system. Thus, it is an interesting problem whether or not the molecular tunneling accompanied by dissociation predicted in the Hubbard model is really realized in a superfluid Fermi gas loaded on an optical lattice. This is also related to the problem about the validity of the Hubbard model for superfluid Fermi gases loaded on optical lattices. 
\par
In this paper, we investigate a bound pair of Fermi atoms in an optical lattice. Including a cosine-shape periodic potential, we calculate molecular wavefunction. We show how the spatial structure of the bound pair changes during the tunneling through the lattice potential. We also compare molecular kinetic properties in the optical lattice potential with results in an attractive Fermi Hubbard model, in order to examine the validity of the Hubbard model for superfluid Fermi gases in optical lattices. 
\par
This paper is organized as follows. In Sec. II, we explain a model optical lattice system, as well as how to calculate molecular wavefunction. Here, we also construct a single-band Hubbard model for the optical lattice system. In Sec. III, we calculate molecular excitations. They are compared with results obtained in the Hubbard model. In Sec. IV, we study the spatial structure of molecular wavefunction from the weak coupling regime to the strong coupling regime. We examine whether or not the virtual dissociation predicted in the Hubbard model occurs in the optical lattice system. In Sec. V, we consider the superfluid phase transition temperature $T_{\rm c}$ in the BEC limit, where the Hubbard model gives $T_{\rm c}=0$. Throughout this paper, we take $\hbar=k_B=1$. We also set the system volume unity.
\par
\section{Bound state in model one-dimensional optical lattice}
\par
We consider two attractively interacting Fermi atoms in a three dimensional system, in the presence of a one-dimensional optical lattice in the $x$-direction. These atoms are assumed to be in different hyperfine states, described by pseudospin $\sigma=\uparrow,\downarrow$. The Hamiltonian is given by\cite{Orso1,Orso2}
\begin{equation}
H=H_0({\bf r}_1)+H_0({\bf r}_2)-U\delta({\bf r}_1-{\bf r}_2),
\label{eq.2.1}
\end{equation}
where $-U$ is the $s$-wave pairing interaction. The one-particle Hamiltonian $H_0$ has the form
\begin{eqnarray}
H_0({\bf r})=-{\nabla^2 \over 2m}
+{E_r s \over 2}
\Bigl(
1-\cos{2\pi x \over d}
\Bigr),
\label{eq.2.2}
\end{eqnarray}
where $m$ is the mass of a Fermi atom. The last term in Eq. (\ref{eq.2.2}) describes the optical lattice in the $x$-direction, the height of which is measured in terms of the atomic recoil energy $E_r\equiv \pi^2/2md^2$. The lattice constant $d$ is related to the wavelength $\lambda$ of laser light as $d=\lambda/2$\cite{Pita}. In this paper, we ignore effects of a trap potential, for simplicity. 
\par
Because of the contact pairing interaction in Eq. (\ref{eq.2.1}), only the singlet pairing is allowed as the spin state of a bound pair. For the spatial part of the molecular wavefunction, noting that Eq. (\ref{eq.2.1}) is periodic in terms of the center of mass coordinate $R \equiv (x_1+x_2)/2$ with the period $d$, one may take
\begin{equation}
\Psi_{\bf q}({\bf r}_1,{\bf r}_2)=\sum_{{\bf p},n_1,n_2}
g_{\bf p}^{n_1,n_2}({\bf q})
\phi^{n_1}_{{\bf p}+{\bf q}/2}({\bf r}_1)
\phi^{n_2}_{-{\bf p}+{\bf q}/2}({\bf r}_2).
\label{eq.2.3}
\end{equation}
Here, $\phi_{\bf p}^n({\bf r})$ is an eigenfunction of the one-particle Hamiltonian $H_0({\bf r})$ in Eq. (\ref{eq.2.2}), with the energy $\varepsilon_{\bf p}^n$, where $n$ is a band index. Since the system is uniform in the $y$- and $z$-direction, the atomic energy has the form, $\varepsilon_{\bf p}^n=\varepsilon_{p_x}^n+(p_y^2+p_z^2)/2m$. Using the Bloch's theorem, we can write the eigenfunction in the form $\phi_{\bf p}^n({\bf r})=e^{i{\bf p}\cdot{\bf r}}u_{p_x}^n(x)$, where $u_{p_x}^n(x)$ is a periodic function satisfying $u_{p_x}^n(x+d)=u_{p_x}^n(x)$. Because of the required antisymmetric property of fermion wavefunction, the spatial part $\Psi_{\bf q}({\bf r}_1,{\bf r}_2)$ must be symmetric with respect to the exchange of ${\bf r}_1$ and ${\bf r}_2$. This is satisfied by imposing the condition $g_{-{\bf p}}^{n_2,n_1}({\bf q})=g_{\bf p}^{n_1,n_2}({\bf q})$ in Eq. (\ref{eq.2.3}).
\par
We note that $g_{\bf p}^{n_1,n_2}({\bf q})$ with $n_1\ne n_2$ describes interband coupling due to the spatial inhomogeneity by the lattice potential. In the extended zone scheme, this means that pairs of two atomic states with ${\bf p}$ and $-{\bf p}+{\bf G}$ contribute to the molecular state when ${\bf q}=0$, where ${\bf G}$ is the reciprocal lattice vector. This is different from the case of a uniform gas, where the molecular wavefunction with ${\bf q}=0$ only involves pairs of atomic states with ${\bf p}$ and $-{\bf p}$ as $\Psi^{\rm uniform}_{{\bf q}=0}({\bf r}_1,{\bf r}_2)=\sum_{\bf p}g_{\bf p} e^{i{\bf p}\cdot{\bf r}_1}e^{-{\bf p}\cdot{\bf r}_2}$.
\par
Substituting Eq. (\ref{eq.2.3}) into the Schr\"odinger equation $H\Psi_{\bf q}=E_{\bf q}\Psi_{\bf q}$, we obtain 
\begin{eqnarray}
g_{\bf p}^{n_1,n_2}({\bf q})
&=&
{U 
\over 
\varepsilon_{{\bf p}+{\bf q}/2}^{n_1}+\varepsilon_{-{\bf p}+{\bf q}/2}^{n_2}-E_{\bf q}
}
\nonumber
\\
&\times&
\sum_{{\bf k},n_3,n_4}
\int d{\bf r}
\phi^{n_1}_{{\bf p}+{\bf q}/2}({\bf r})^*
\phi^{n_2}_{-{\bf p}+{\bf q}/2}({\bf r})^*
\phi^{n_3}_{{\bf k}+{\bf q}/2}({\bf r})
\phi^{n_4}_{-{\bf k}+{\bf q}/2}({\bf r})
g_{\bf k}^{n_3,n_4}({\bf q}).
\label{eq.2.4}
\end{eqnarray}
Since we are interested in the pair tunneling through the lattice potential, we take ${\bf q}=(q,0,0)$ in this paper. In addition, to examine the spatial structure of the molecular wavefunction $\Psi_{\bf q}({\bf r}_1,{\bf r}_2)$ in the $x$-direction, we set $y_1=y_2\equiv y$ and $z_1=z_2\equiv z$ in Eq. (\ref{eq.2.3}). The resulting molecular wavefunction $\Psi_{\bf q}(x_1,x_2)\equiv\Psi_{\bf q}(x_1,y,z;x_2,y,z)$ does not depend on $y$ and $z$. Introducing the relative coordinate $r\equiv x_1-x_2$ and the center of mass coordinate $R=(x_1+x_2)/2$, one can rewrite $\Psi_{\bf q}(x_1,x_2)$ in the form
\begin{eqnarray}
\Psi_{\bf q}(r,R)
&\equiv&\Psi_{\bf q}(R+r/2,R-r/2)
\nonumber
\\
&=&
e^{iqR}\sum_{p_x,n_1,n_2}f_{p_x}^{n_1,n_2}({\bf q})
u_{p_x+q/2}^{n_1}(R+r/2)u_{-p_x+q/2}^{n_2}(R-r/2).
\label{eq.2.5}
\end{eqnarray} 
The coefficient $f_{p_x}^{n_1,n_2}({\bf q})\equiv\sum_{p_y,p_z}g_{\bf p}^{n_1,n_2}({\bf q})$ obeys the equation
\begin{eqnarray}
f_{p_x}^{n_1,n_2}({\bf q})
&=&
\sum_{p_y,p_z}{U 
\over 
\varepsilon_{{\bf p}+{\bf q}/2}^{n_1}+\varepsilon_{-{\bf p}+{\bf q}/2}^{n_2}-E_{\bf q}
}
\nonumber
\\
&\times&
\sum_{k_x,n_3,n_4}
{1 \over d}\int_0^d dx
u^{n_1}_{p_x+q/2}(x)^*
u^{n_2}_{-p_x+q/2}(x)^*
u^{n_3}_{k_x+q/2}(x)
u^{n_4}_{-k_x+q/2}(x)
f_{k_x}^{n_3,n_4}({\bf q}).
\label{eq.2.6}
\end{eqnarray}
In the cosine-shape periodic potential, $u_{p_x}^n(x)$ may be written as
\begin{equation}
u_{p_x}^n(x)=\sum_l C^n_{p_x}(l) e^{i{2\pi l \over d}x},
\label{eq.2.7}
\end{equation}
where $C_{p_x}^n(l)$ is determined by the equation
\begin{equation}
\Bigl[
{p_x^2 \over 2m}+{E_rs \over 2}-\varepsilon_{p_x}^n
\Bigr]C_{p_x}^n(l)
-{E_rs \over 4}[C_{p_x}^n(l+1)+C_{p_x}^n(l-1)]=0.
\label{eq.2.8}
\end{equation}
Substituting Eq. (\ref{eq.2.7}) into Eq. (\ref{eq.2.6}), we execute the integration over $x$. Then we obtain
\begin{eqnarray}
f_{p_x}^{n_1,n_2}({\bf q})
=
\sum_{p_y,p_z}{U 
\over 
\varepsilon_{{\bf p}+{\bf q}/2}^{n_1}+\varepsilon_{-{\bf p}+{\bf q}/2}^{n_2}-E
}
\sum_{k_x,n_3,n_4}
\eta_{n_3,n_4}^{n_1,n_2}({\bf q};p_x,k_x)f_{k_x}^{n_3,n_4}({\bf q}),
\label{eq.2.9}
\end{eqnarray}
where
\begin{equation}
\eta_{n_3,n_4}^{n_1,n_2}({\bf q};p_x,k_x)=\sum_{l_1,l_2,l_3,l_4 \atop l_1+l_2=l_3+l_4}
C_{p_x+q/2}^{n_1}(l_1)
C_{-p_x+q/2}^{n_2}(l_2)
C_{k_x+q/2}^{n_3}(l_3)
C_{-k_x+q/2}^{n_4}(l_4).
\label{eq.2.10}
\end{equation}
\par
The contact interaction in Eq. (\ref{eq.2.1}) brings about the ultraviolet divergence in Eq. (\ref{eq.2.9}). Thus, we introduce the cutoff $\Lambda_c$ in the summations over $(k_x,n_3,n_4)$, as well as the cutoff $\Lambda_{\perp}$ in the summation over $p_\perp\equiv\sqrt{p_y^2+p_z^2}$ in Eq. (\ref{eq.2.9}). As usual, we eliminate effects of these momentum cutoffs by regularizing Eq. (\ref{eq.2.9}), which is achieved by introducing the two-body scattering length $a_s$ given by
\begin{eqnarray}
{4\pi a_s \over m}=-
{U \over 1-\alpha U}.
\label{eq.2.11}
\end{eqnarray}
Here, $\alpha$ has the form
\begin{equation}
\alpha=\sum_{p_y,p_z}^{\Lambda_\perp}\sum_{p_x}^{\Lambda_c}{m\over p^2}
={m\Lambda_\perp \over 4\pi^2}
\Bigl[
{\Lambda_c \over \Lambda_\perp}
\ln{\Lambda_c^2+\Lambda_\perp^2 \over \Lambda_c^2}
+2\tan^{-1}{\Lambda_c \over \Lambda_\perp}
\Bigr].
\label{eq.2.12}
\end{equation} 
When we take $\Lambda_c/\Lambda_\perp\gg 1$, Eq. (\ref{eq.2.11}) reduces to the familiar expression, 
\begin{equation}
{4\pi a_s \over m}
=-{U \over \displaystyle 1-U\sum_{\bf p}^{\Lambda_c}{m \over p^2}}
=-{U \over \displaystyle 1-U{m\Lambda_c \over 2\pi^2}}.
\label{eq.2.13}
\end{equation}
\par
Executing the summations over $p_y$ and $p_z$ in Eq. (\ref{eq.2.9}), one finds
\begin{eqnarray}
f_{p_x}^{n_1,n_2}({\bf q})
=
\Gamma_{\bf q}^{n_1,n_2}(p_x)
\sum_{k_x,n_3,n_4}^{\Lambda_c}
\eta_{n_3,n_4}^{n_1,n_2}({\bf q};p_x,k_x)f_{k_x}^{n_3,n_4}({\bf q}),
\label{eq.2.14}
\end{eqnarray}
where
\begin{equation}
\Gamma_{\bf q}^{n_1,n_2}(p_x)=
{mU \over 4\pi}
\ln
{(\varepsilon_{p_x+q/2}^{n_1}+\varepsilon_{-p_x+q/2}^{n_2})-E_{\bf q}+{\Lambda_\perp^2 \over m}
\over 
(\varepsilon_{p_x+q/2}^{n_1}+\varepsilon_{-p_x+q/2}^{n_2})-E_{\bf q}}.
\label{eq.2.15}
\end{equation}
\par
We numerically solve the eigenvalue equation (\ref{eq.2.14}) to determine the molecular excitation spectrum $E_{\bf q}$, as well as $f_{p_x}^{n_1,n_2}({\bf q})$. The molecular wavefunction $\Psi_{\bf q}(r,R)$ is calculated from Eq. (\ref{eq.2.5}). Since the current experiments on superfluid lattice Fermi gas are using a weak optical lattice potential\cite{MIT1,MIT2}, we take $s=3$. For the momentum cutoffs $\Lambda_c$ and $\Lambda_\perp$, it is difficult to take very large values because of computational problem. In this paper, we choose the value of $\Lambda_c$ so as to be able to include energy bands up to $n=14$. For the cutoff $\Lambda_\perp$, we set $\Lambda_\perp=\Lambda_c/6$. Although numerical results on $E_{\bf q}$ still weakly depend on $\Lambda_c$ and $\Lambda_\perp$, we expect that the essence of our results would be unaltered even when larger values of $\Lambda_c$ and $\Lambda_\perp$ are used. We also find from numerical results that the truncation of the summations over $(p_x,n_1,n_2)$ at $\Lambda_c$ in Eq. (\ref{eq.2.5}) affects the spatial structure of the molecular wavefunction around $r=0$. For this problem, using the fact that one-particle wavefunction $\phi_{\bf p}^n({\bf r})$ reduces to the plane wave in the high energy limit, we take into account the contribution from higher momentum region than $\Lambda_c$ by approximating $\phi_{\bf p}^n({\bf r})$ to the plane wave in calculating $\Psi_{\bf q}(r,R)$. Although this prescription cannot completely eliminate  cutoff effects on $\Psi_{\bf q}(r\sim 0,R)$, we can still study interesting molecular tunneling properties, using the spatial structure of $\Psi_{\bf q}(r,R)$.
\par

%%%%%%%%%%%%%%%%%%%%%%%%%%%%%%%%%%%%%%%%%%%%%%%%%%%%%%%%%%%%%%%%%%%%%%%%%%%%%%%
\begin{figure}[t]
\includegraphics[width=8cm,height=5.5cm]{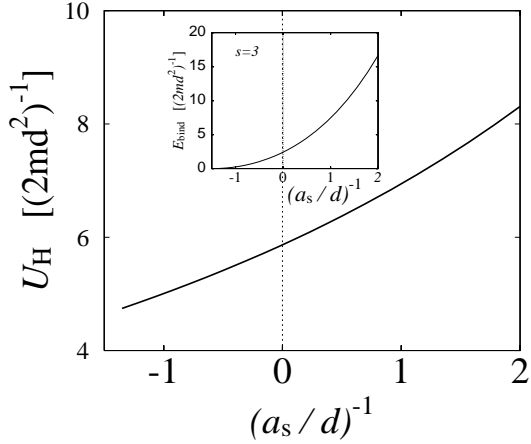}
\caption{
Calculated interaction $U_H$ in the effective Hubbard model in Eq. (\ref{eq.3.0}), as a function of the pairing interaction measured in terms of the inverse scattering length $a_s$. The inset shows the molecular binding energy $E_{\rm bind}$ obtained from Eq. (\ref{eq.2.14}). 
\label{fig1}
}
\end{figure}
%%%%%%%%%%%%%%%%%%%%%%%%%%%%%%%%%%%%%%%%%%%%%%%%%%%%%%%%%%%%%%%%%%%%%%%%%%%%%%%

Besides the pair wavefunction $\Psi_{\bf q}(r,R)$, we also consider a single-band Fermi Hubbard model for the present periodic potential model in Eq. (\ref{eq.2.1}). The single-band Hubbard model in momentum space is given by
\begin{equation}
H=\sum_{{\bf p},\sigma}\varepsilon_{\bf p}
c^\dagger_{{\bf p}\sigma}c_{{\bf p}\sigma}
-{U_H \over N}\sum_{{\bf p},{\bf p}',{\bf q}}
c^\dagger_{{\bf p}+{\bf q}/2\uparrow}c^\dagger_{-{\bf p}+{\bf q}/2\downarrow}c_{-{\bf p}'+{\bf q}/2\downarrow}c_{{\bf p}'+{\bf q}/2\uparrow},
\label{eq.3.0}
\end{equation}
where $N$ is the total number of lattice sites in the $x$-direction. $c_{{\bf p}\sigma}^\dagger$ is the creation operator of a Fermi atom with pseudospin $\sigma=\uparrow,\downarrow$. In Eq. (\ref{eq.3.0}), we take the band dispersion $\varepsilon_{\bf p}$ so as to be equal to the lowest energy band $\varepsilon_{\bf p}^{n=1}$ calculated in the periodic potential system given by Eq. (\ref{eq.2.2})\cite{note}. The attractive interaction $-U_H$ is taken so that Eq. (\ref{eq.3.0}) can reproduce the molecular binding energy $E_{\rm bind}\equiv |E_{{\bf q}=0}|$ obtained in the original periodic potential model given by Eq. (\ref{eq.2.1}). Setting the molecular state as $|\Psi_H\rangle=\sum_{\bf p}{\tilde g}_{\bf p}c^\dagger_{{\bf p}+{\bf q}/2\uparrow}c^\dagger_{-{\bf p}+{\bf q}/2\downarrow}|0\rangle$, we obtain the equation for the energy $E_{\bf q}^H$ of a bound state as
\begin{equation}
1={U_H \over N}\sum_{\bf p}{1 \over \varepsilon_{{\bf p}+{\bf q}/2}+\varepsilon_{-{\bf p}+{\bf q}/2}-E_{\bf q}^H}.
\label{eq.3.1}
\end{equation}
Thus, $U_H$ is given by
\begin{equation}
U_H^{-1}={1 \over N}\sum_{\bf p}{1 \over 2\varepsilon_{\bf p}+E_{\rm bind}}.
\label{eq.3.2}
\end{equation} 
\par
Figure \ref{fig1} shows the calculated Hubbard interaction $U_H$. In obtaining this result, we have used the binding energy $E_{\rm bind}$ obtained from Eq. (\ref{eq.2.14}), which is shown in the inset of Fig.\ref{fig1}. We briefly note that, in a periodic potential, a two-body bound state is possible even for negative scattering length\cite{Orso1,Zoller}. (See the inset in Fig.\ref{fig1}.) In a uniform system, a two-body bound state is possible only when $a_s^{-1}>0$.   
\par

\section{Excitations and band mass of a bound pair of Fermi atoms}

%%%%%%%%%%%%%%%%%%%%%%%%%%%%%%%%%%%%%%%%%%%%%%%%%%%%%%%%%%%%%%%%%%%%%%%%%%%%%%%
\begin{figure}[t]
\includegraphics[width=7cm,height=13cm]{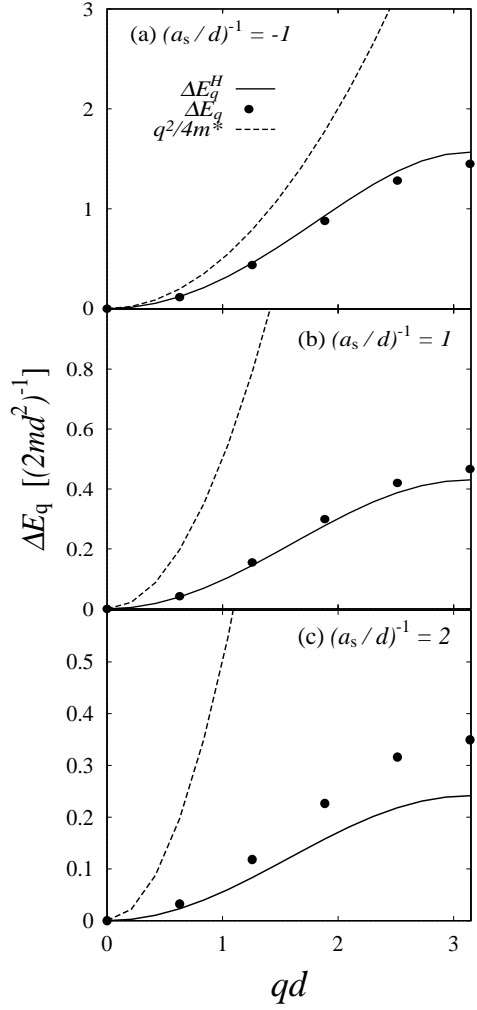}
\caption{
Molecular excitation spectrum $\Delta E_{\bf q}\equiv E_{\bf q}-E_{{\bf q}=0}$. We take ${\bf q}=(q,0,0)$ and $s=3$.  $\Delta E_{\bf q}^H\equiv E^H_{\bf q}-E^H_{{\bf q}=0}$ is the result obtained in the Hubbard model. $q^2/4m^*$ is the molecular kinetic energy, when we assume a uniform system except that the atomic bare mass $m$ is replaced by the band mass of the lowest atomic band, given by $m^*=({\partial^2 \varepsilon_{\bf p}/\partial p_x^2})_{{\bf p}\to 0}^{-1}$.
\label{fig2}
}
\end{figure}
%%%%%%%%%%%%%%%%%%%%%%%%%%%%%%%%%%%%%%%%%%%%%%%%%%%%%%%%%%%%%%%%%%%%%%%%%%%%%%%%
Figure \ref{fig2} shows the molecular excitation spectrum $\Delta E_{\bf q}\equiv E_{\bf q}-E_{{\bf q}=0}$ (${\bf q}=(q,0,0)$). When the pairing interaction is not strong (panels (a) and (b)), we find that the Hubbard model in Eq. (\ref{eq.3.0}) well reproduces the excitation spectrum $\Delta E_{\bf q}$ obtained in the periodic potential system. On the other hand, when $(a_s/d)^{-1}=2$ (panel(c)), the Hubbard model underestimates $\Delta E_{\bf q}$.
\par
\begin{figure}[t]
\includegraphics[width=8cm,height=5.5cm]{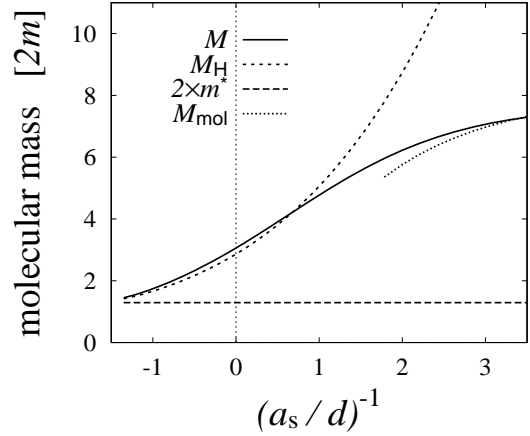}
\caption{
Molecular band mass $M=({\partial^2 E_{\bf q}/\partial q_x^2})^{-1}_{{\bf q}\to 0}$, as a function of the inverse scattering length $a_s$. $M_H$ shows the result obtained in the Hubbard model, given by $M_H=({\partial^2 E^H_{\bf q}/\partial q_x^2})^{-1}_{{\bf q}\to 0}$, where $E_{\bf q}^H$ is determined by Eq. (\ref{eq.3.1}). $M_{\rm mol}$ is the molecular mass obtained from Eq. (\ref{eq.3.5}).
\label{fig3}
}
\end{figure}

Figure \ref{fig3} shows the molecular (band) mass $M$ in the $x$-direction, defined by
\begin{equation}
M\equiv
\Bigl(
{\partial^2 E_{\bf q} \over \partial q_x^2}
\Bigr)^{-1}_{{\bf q}\to 0}.
\label{eq.3.3}
\end{equation}
In the weak-coupling regime ($(a_s/d)^{-1}\simeq -1$), the magnitude of the molecular mass $M$ is close to twice the atomic band mass, given by $m^*=(\partial^2\varepsilon_{\bf p}/\partial p_x^2)^{-1}_{{\bf p}\to 0}$. This can be also seen in Fig.\ref{fig2}(a), where the excitation spectrum $\Delta E_{\bf q}$ is well approximated to $q^2/4m^*$ in the small momentum region. When $(a_s/d)^{-1}\sim -1$, since the molecule is weakly binding, the molecular motion is dominantly determined by the sum of two atomic band motions, which leads to $M\simeq 2m^*$. 
\par
As one increases the strength of the pairing interaction, Fig.\ref{fig3} shows that the molecular band mass $M$ becomes heavier than $2m^*$. (See also Figs.\ref{fig2}(b) and 2(c).) This mass enhancement can be described by the Hubbard model ($M_H$ in Fig.\ref{fig3}) when $(a_s/d)^{-1}\lesssim 1$. However, when $(a_s/d)^{-1}\gesim 1$, the Hubbard model overestimates the molecular mass. In the strong coupling limit, while $M$ approaches a constant value\cite{Orso1}, $M_H$ diverges\cite{Nozieres}. 
\par

%%%%%%%%%%%%%%%%%%%%%%%%%%%%%%%%%%%%%%%%%%%%%%%%%%%%%%%%%%%%%%%%%%%%%%%%%%%%%%
%%%%%%%%%%%%%%%%%%%%%%%%%%%%%%%%%%%%%%%%%%%%%%%%%%%%%%%%%%%%%%%%%%%%%%%%%%%%%%%

\begin{figure}
\includegraphics[width=8cm,height=5.5cm]{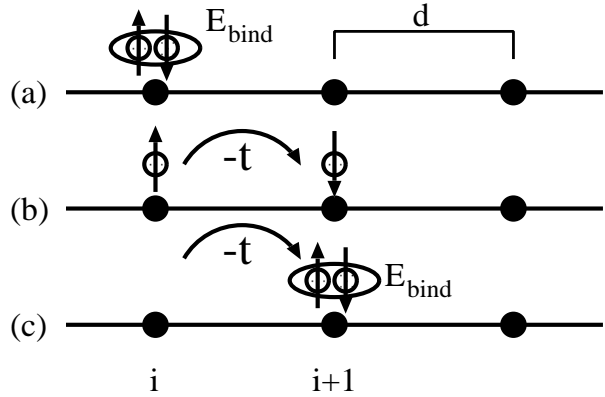}
\caption{
Simple model of pair tunneling in the strong coupling regime of the tight-binding model with the nearest-neighbor atomic hopping $-t$. Solid and open circles represent lattice sites and atoms, respectively. When a molecule moves from the $i$-th site to the $(i+1)$-th site, each atom has to move one by one. As a result, the modulate dissociates into two atoms in the intermediate state (panel (b)), so that the energy in the intermediate state is higher than that in the initial state by the binding energy $E_{\rm bind}$. This tunneling process leads to the molecular hopping matrix element $-t_M=2t^2/E_{\rm bind}$.
\label{fig4}
}
\end{figure}

The difference between $M$ and $M_H$ in the strong coupling regime originates from different tunneling mechanisms between the original optical lattice system and the effective Hubbard model. As mentioned in the introduction, molecular motion in the Hubbard model is accompanied by virtual dissociation in the strong coupling regime\cite{Nozieres}. To see this in a simple manner, we consider the model shown in Fig.\ref{fig4}, where a tightly bound molecule with the binding energy $E_{\rm bind}$ moves from the $i$-th site to the $(i+1)$-th site. Noting that the creation of the intermediate state in Fig.\ref{fig4}(b) costs $E_{\rm bind}$, we obtain the nearest-neighbor molecular hopping matrix element as $-t_M=-2t^2/E_{\rm bind}$\cite{Nozieres}, where $-t$ is the nearest-neighbor atomic transfer matrix element. When we only retain the tunneling process shown in Fig.\ref{fig4} by assuming a small $t$, we obtain the molecular band $\varepsilon^M_{q_x}=-2t_M\cos (q_xd)$, giving the  molecular band mass ${\tilde M}=E_{\rm bind}/(2td)^2$. Because $E_{\rm bind}\to\infty$ in the strong-coupling limit, ${\tilde M}$ diverges. Although this is a simple evaluation, the enhancement of $M_H$ in the strong coupling regime shown in Fig.\ref{fig3} is found to be directly related to the molecular tunneling accompanied by virtual dissociation. 
\par

%%%%%%%%%%%%%%%%%%%%%%%%%%%%%%%%%%%%%%%%%%%%%%%%%%%%%%%%%%%%%%%%%%%%%%%%%%%%%%%
%%%%%%%%%%%%%%%%%%%%%%%%%%%%%%%%%%%%%%%%%%%%%%%%%%%%%%%%%%%%%%%%%%%%%%%%%%%%%%%

On the other hand, Orso and co-workers\cite{Orso1} showed that the molecular tunneling mechanism in the strong coupling regime of cosine-shape periodic potential system is quite different. In this regime, since the molecular size is much smaller than the lattice spacing $d$, the $r$-dependence of the molecular wavefunction $\Psi_{\bf q}(r,R)$ is close to that in a uniform system, given by 
\begin{equation}
\Psi({\bar r})={1 \over \sqrt{2\pi a_s}}{1 \over {\bar r}}e^{-{\bar r}/a_s},
\label{eq.3.4}
\end{equation}
where ${\bar r}=\sqrt{x^2+y^2+z^2}$. When we extract terms depending on the center of mass coordinate $R$ from Eq. (\ref{eq.2.1}), we obtain\cite{Orso1}
\begin{eqnarray}
H_R
&=&
-{1 \over 4m}{\partial^2 \over \partial R^2}-E_ss\cos{\pi r \over d}\cos{2\pi R \over d}
\nonumber
\\
&\simeq&
-{1 \over 4m}{\partial^2 \over \partial R^2}-{2dE_ss \over \pi a_s}
\tan^{-1}{\pi a_s \over 2d}\cos{2\pi R \over d}.
\label{eq.3.5}
\end{eqnarray}
In obtaining the last expression, we have replaced the factor $\cos({\pi r/d})$ by the expectation value $\langle\Psi({\bar r})|\cos(\pi r/d)|\Psi({\bar r})\rangle=(2d/\pi a_s)\tan^{-1}(\pi a_s/2d)$. In the strong-coupling limit ($a_s^{-1}\to+\infty$), Eq. (\ref{eq.3.5}) reduces to
\begin{equation}
H_R=-{1 \over 4m}{\partial^2 \over \partial R^2}-E_ss\cos{2\pi R \over d}.
\label{eq.3.6}
\end{equation}
Equation (\ref{eq.3.6}) shows that a molecule feels the periodic potential with the {\it finite} height $2E_ss$ in the strong coupling limit\cite{Orso1}. Namely, the molecular band mass $M$ remains finite, in contrast to the case of Hubbard model. As shown in Fig.\ref{fig3}, the molecular band mass $M_{\rm mol}$ calculated from Eq. (\ref{eq.3.5}) explains the behavior of $M$ in the strong coupling regime. This means that the bound pair moves {\it without} dissociation in the strong coupling regime of optical lattice system.
\par
Although the molecular tunneling accompanied by the dissociation does not occur in the strong coupling regime of the optical lattice system, we can still expect the possibility of this interesting tunneling phenomenon somewhere in the BCS-BEC crossover region, especially in the intermediate coupling regime ($(a_s/d)^{-1}\sim 0$), because the enhancement of $M$ in this regime is in good agreement with $M_H$, as shown in Fig.\ref{fig3}. In the next section, we explore this possibility, based on the analysis of molecular wavefunction.

%%%%%%%%%%%%%%%%%%%%%%%%%%%%%%%%%%%%%%%%%%%%%%%%%%%%%%%%%%%%%%%%%%%%%%%%%%%%%%%
\begin{figure}
\includegraphics[width=16cm,height=13cm]{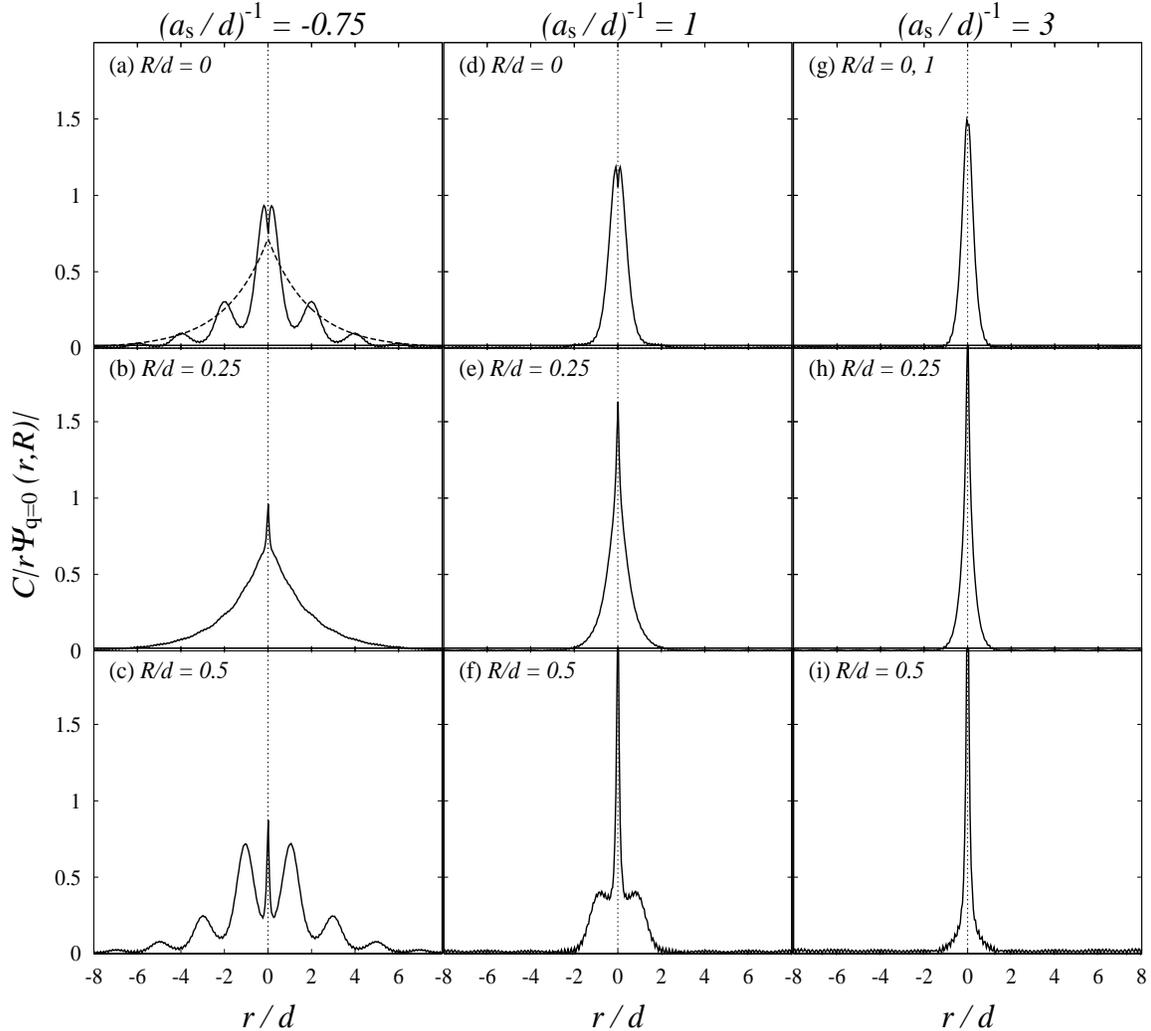}
\caption{Calculated molecular wavefunction $\Psi_{{\bf q}=0}(r,R)$ as a function of the relative coordinate $r=x_1-x_2$. In this figure, we show $|r\Psi_{{\bf q}=0}(r,R)|$ to eliminate the $r^{-1}$ behavior, which also appears in the molecular wavefunction in a uniform system. The coefficient $C$ is chosen as $C^{-1}=\sqrt{\int_{-\infty}^\infty dr|r\Psi_{\bf q}(r,R)|^2}$. The dashed line in panel (a) shows ${\tilde \Psi}_{{\bf q}=0}(r)$ in Eq. (\ref{eq.4.1}). Sharp peaks and dips seen around $r=0$ may be artifacts, originating from the finite cutoff $\Lambda_c$ used in numerical calculations. (See the text.) 
\label{fig5}
}
\end{figure}
%%%%%%%%%%%%%%%%%%%%%%%%%%%%%%%%%%%%%%%%%%%%%%%%%%%%%%%%%%%%%%%%%%%%%%%%%%%%%%%

%%%%%%%%%%%%%%%%%%%%%%%%%%%%%%%%%%%%%%%%%%%%%%%%%%%%%%%%%%%%%%%%%%%%%%%%%%%%%%%
\begin{figure}
\includegraphics[width=8cm,height=4cm]{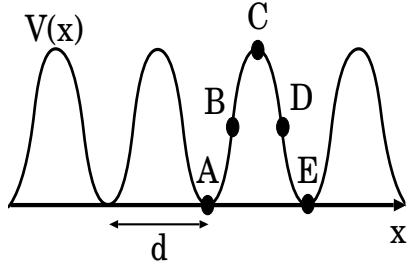}
\caption{Center of mass position of the molecular wavefunction $\Psi_{\bf q}(r,R)$ in Fig.\ref{fig5}. $V(x)=-(E_ss/2)[\cos(2\pi x/d)-1]$ is the periodic potential. A: $R/d=0$. B: $R/d=0.25$. C: $R/d=0.5$. Results for D ($R/d=0.75$) and E ($R/d=1$) are the same as those at $R/d=0.25$ and $R/d=0$, respectively.
\label{fig6}
}
\end{figure}
%%%%%%%%%%%%%%%%%%%%%%%%%%%%%%%%%%%%%%%%%%%%%%%%%%%%%%%%%%%%%%%%%%%%%%%%%%%%%%%

%%%%%%%%%%%%%%%%%%%%%%%%%%%%%%%%%%%%%%%%%%%%%%%%%%%%%%%%%%%%%%%%%%%%%%%%%%%%%%%
\begin{figure}[t]
\includegraphics[width=8cm,height=5.5cm]{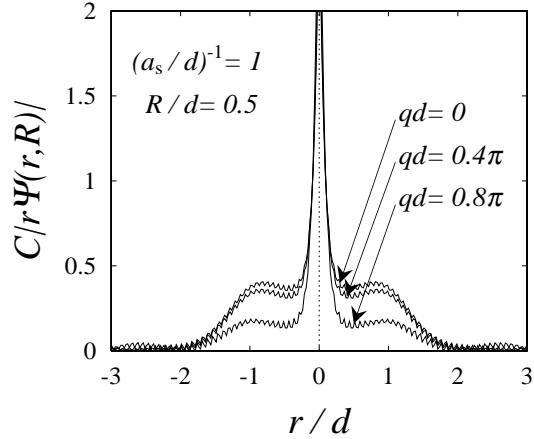}
\caption{Spatial variation of the molecular wavefunction $\Psi_{\bf q}(r,R)$ in the case of finite ${\bf q}=(q,0,0)$. 
\label{fig7}
}
\end{figure}
%%%%%%%%%%%%%%%%%%%%%%%%%%%%%%%%%%%%%%%%%%%%%%%%%%%%%%%%%%%%%%%%%%%%%%%%%%%%%%%
\par
\section{Molecular wavefunction and virtual dissociation in optical lattice}
\par
Figure \ref{fig5} shows the calculated molecular wavefunction $\Psi_{{\bf q}=0}(r,R)$, as a function of the relative coordinate $r=x_1-x_2$. In this figure, the upper and lower panels show the cases when the molecular center of mass position is at the bottom of the periodic potential $R=0$ and at the top of the potential $R/d=0.5$, respectively ((A) and (C) in Fig.\ref{fig6}, respectively). The middle panels show the case of $R/d=0.25$ (and also $R/d=0.75$) ((B) and (D) in Fig.\ref{fig6}). Namely, Fig. \ref{fig5} shows how the spatial structure of the molecule varies when the molecule moves from (A) to (E) in Fig.\ref{fig6}.
\par
When the pairing interaction is weak (Figs.\ref{fig5}(a)-(c)), the wavefunction spreads out. Except for panel (b), one finds oscillating structures, originating from the presence of optical lattice potential. For a given binding energy $E_{\rm bind}$ and atomic band mass $m^*$, when we calculate the molecular wavefunction ignoring other lattice effects, we obtain
\begin{equation}
{\tilde \Psi}_{{\bf q}=0}(r)={{\tilde C} \over |r|}
e^{-\sqrt{m^*E_{\rm bind}}|r|},
\label{eq.4.1}
\end{equation}
where ${\tilde C}$ is a normalization constant\cite{notez}. Apart from the oscillating structure, the overall spatial variation of $\Psi_{{\bf q}=0}(r)$ can be described by Eq. (\ref{eq.4.1}), as shown in Fig.\ref{fig5}(a). This means that the molecular size in this regime is dominated by the binding energy $E_{\rm bind}$ and atomic band mass $m^*$ appearing in Eq. (\ref{eq.4.1}).
\par
Comparing Fig.\ref{fig5}(a) with Fig.\ref{fig5}(c), we find that the peak positions in $\Psi_{{\bf q}=0}(r)$ are different between the two. In panel (a), in addition to the central peak at $r=0$, satellite peaks can be seen at $r/d=\pm 2,\pm 4,\cdot\cdot\cdot$. In panel (c), the satellite peaks appear at $r/d=\pm 1,\pm 3,\cdot\cdot\cdot$. In the former case, to satisfy $R=0$ avoiding the potential energy, one should put two atoms at $(x_1,x_2)=(0,0),~(\pm d,\mp d),~(\pm 2d,\mp 2d),\cdot\cdot\cdot$. In the relative coordinate $r=x_1-x_2$, these configurations immediately explain the peak positions in Fig.\ref{fig5}(a). In the same way, the configurations which satisfy $R/d=0.5$ and avoid the potential energy loss are $(x_1,x_2)=(0,d),~(d,0),~(-d,2d),\cdot\cdot\cdot$. These again explain the peak positions in Fig.\ref{fig5}(c). The case of $R/d=0.25$ is considered to involve both configurations, so that the oscillating structure is cancelled out to disappear, as shown in Fig.\ref{fig5}(b).
\par
As one increases the strength of the pairing interaction, the molecular wavefunction shrinks, reflecting the increase of the binding energy $E_{\rm bind}$. In the intermediate coupling regime shown in Figs.\ref{fig5}(d)-(f), while no satellite peak can be seen in panels (d) and (e), one finds two satellite peaks at $r=\pm d$ in panel (f). This means that the bound pair partially dissociates into two atoms at $x=0$ and $x=d$, when the center of mass position is at the top of the lattice potential $R/d=0.5$. As shown in Fig.\ref{fig7}, these satellite peaks at $r=\pm d$ also exist when the molecule has a finite momentum $q$ in the $x$-direction. Thus, we find that the pair tunneling accompanied by dissociation predicted in the Hubbard model really occurs in the {\it intermediate} coupling regime of optical lattice system.
\par
However, in the strong coupling regime shown in Figs.\ref{fig5}(g)-(i), the satellite peaks are absent even when the molecular center of mass position is at $R/d=0.5$ (panel (i)). Namely, the molecule tunnels through the lattice potential without dissociation in this regime, which is consistent with the discussion in the previous section.
\par
\section{Superfluid phase transition temperature in the strong-coupling limit of lattice Fermi gas}
\par
Since the pair mass $M$ in the optical lattice system actually remains finite in the strong coupling limit, we can expect a finite superfluid phase transition temperature $T_{\rm c}$ even in the BEC limit, where the Hubbard model predicts the vanishing $T_{\rm c}$\cite{Tamaki,Micnus,Levin2}. In this section, we evaluate $T_{\rm c}$ in the BEC limit, including the finite value of $M$.
\par

%%%%%%%%%%%%%%%%%%%%%%%%%%%%%%%%%%%%%%%%%%%%%%%%%%%%%%%%%%%%%%%%%%%%%%%%%%%%%%%
\begin{figure}[t]
\includegraphics[width=8cm,height=5.5cm]{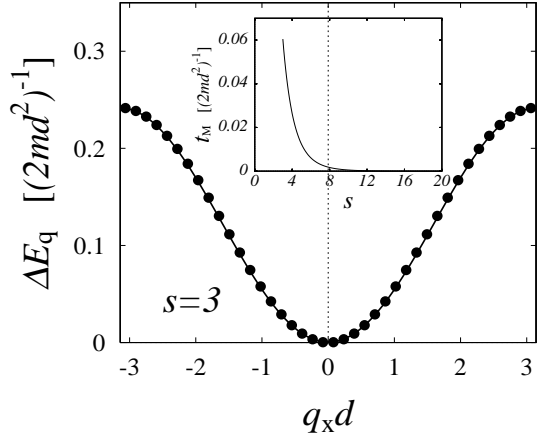}
\caption{Dispersion of molecular excitation spectrum $\Delta E_{\bf q}=E_{\bf q}-E_{q=0}$ in the $q_x$-direction in the BEC limit (solid line). We take $s=3$ and ${\bf q}=(q_x,0,0)$. This result is obtained from the model given by the Hamiltonian in Eq. (\ref{eq.5.2}). Solid circles show the result in the tight-binding model given by the first term in Eq. (\ref{eq.5.1}), where $t_M$ shown in the inset is used. $t_M$ is determined so that the band width $12t_M$ of the Bose Hubbard model in Eq. (\ref{eq.5.1}) can be equal to that of the lowest energy band obtained from the model Hamiltonian in Eq. (\ref{eq.5.2}).
\label{fig8}
}
\end{figure}
%%%%%%%%%%%%%%%%%%%%%%%%%%%%%%%%%%%%%%%%%%%%%%%%%%%%%%%%%%%%%%%%%%%%%%%%%%%%%%%

For this purpose, we consider the BEC limit of a two-component Fermi gas in a three-dimensional cubic optical lattice.  The Hamiltonian is given by
\begin{equation}
H=\sum_{j}H_0({\bf r}_j)-U\sum_{i<j}\delta({\bf r}_i-{\bf r}_j),
\label{eq.5.0}
\end{equation}
where the one-particle Hamiltonian $H_0({\bf r})$ has the form
\begin{eqnarray}
H_0({\bf r})=-{\nabla^2 \over 2m}
+{E_s s \over 2}
\Bigl(
3
-\cos{2\pi x \over d}-\cos{2\pi y \over d}-\cos{2\pi z \over d}
\Bigr).
\label{eq.5.1b}
\end{eqnarray}
In the BEC limit, Cooper pairs have been already formed above $T_{\rm c}$ and the pair size is much smaller than the lattice constant $d$. In this case, one can treat the Cooper pairs as point bosons. Thus, instead of Eq. (\ref{eq.5.0}), one may consider the simpler Hamiltonian, 
\begin{equation}
{\bar H}\equiv H_{R_x}+H_{R_y}+H_{R_z},
\label{eq.5.2}
\end{equation}
where $H_R$ is given by Eq. (\ref{eq.3.6}). In addition, when the lattice potential is strong, the motion of bosons can be described by the tight-binding model with the nearest-neighbor boson hopping $-t_M$. Indeed, as shown in Fig.\ref{fig8}, the molecular excitation spectrum obtained from the model Hamiltonian in Eq. (\ref{eq.5.2}) is well approximated to the tight-binding dispersion $E_{\bf q}^M=-2t_M\sum_{j=x,y,z}[\cos q_jd-1]$ when $s\gesim 3$. We also note that, when the band gap between $\varepsilon_{\bf p}^{n=1}$ and $\varepsilon_{\bf p}^{n=2}$ in the original fermion system is very large in a strong lattice potential, one can ignore multiple occupation of bosons. These situations can be conveniently described by the Bose Hubbard model with infinitely large on-site repulsive interaction $U_M\to+\infty$,
\begin{equation}
H_M=-t_M\sum_{(i,j)}[b_i^\dagger b_j+h.c.]
+{U_M \over 2}\sum_i n_i^M(n_i^M-1),
\label{eq.5.1}
\end{equation}
where $b_i^\dagger$ is the creation operator of a (molecular) boson, and $n_i^M\equiv b_i^\dagger b_i$. The first term describes the boson hopping between nearest-neighbor sites, where the summation is taken over the nearest-neighbor pairs.
\par
When the original fermion system is in the half-filling, the corresponding boson density equals $n_M=0.5$ per lattice site. In this case, when we describe the occupied site and vacant site by pseudospin $\uparrow$ and $\downarrow$, respectively, the Bose Hubbard model with $U_M\to+\infty$ in Eq. (\ref{eq.5.1}) can be mapped onto the three-dimensional XY-model, 
\begin{eqnarray}
H_{XY}
&=&t_M\sum_{(i,j)}(S_+^iS_-^j+S_+^jS_-^i)
\nonumber
\\
&=&2t_M\sum_{(i,j)}(S_x^iS_x^j+S_y^iS_y^j).
\label{eq.5.3}
\end{eqnarray}
Here, $S_\pm^i=S_x^i\pm iS_y^i$, and $S_x^i$ and $S_j^i$ are $S=1/2$ spin operators. Evaluating the phase transition temperature $T_{\rm c}$ of this spin system within the simple mean-field approximation, one finds
\begin{equation}
T_{\rm c}=3t_M.
\label{eq.5.4}
\end{equation}
We note that this result can be also obtained from Eq. (\ref{eq.5.1}) without mapping onto the spin model. We explain the outline of this alternative derivation in the Appendix.
\par
Figure \ref{fig9} shows the calculated $T_{\rm c}$ in the strong coupling BEC limit of a superfluid Fermi gas loaded on the cubic optical lattice (half-filling case). Comparing this result with the maximum $T_{\rm c}\sim0.04\varepsilon_{\rm F}$ obtained in the intermediate coupling regime of the Fermi Hubbard model\cite{Tamaki,Micnus}, we find that, although $T_{\rm c}$ in the BEC limit remains finite due to the finite magnitude of $M$, it is still low when the lattice potential is strong ($s\gg 1$).

%%%%%%%%%%%%%%%%%%%%%%%%%%%%%%%%%%%%%%%%%%%%%%%%%%%%%%%%%%%%%%%%%%%%%%%%%%%%%%%
\begin{figure}[t]
\includegraphics[width=8cm,height=5.5cm]{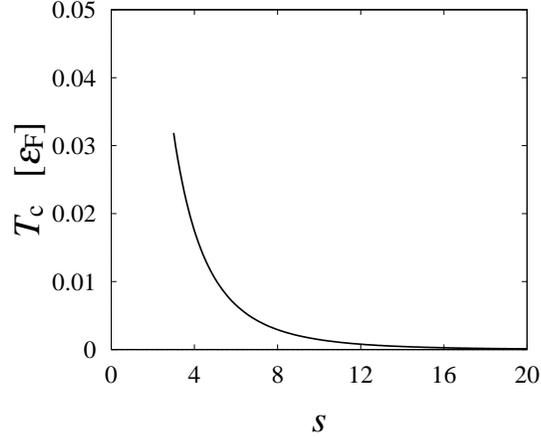}
\caption{Superfluid phase transition temperature $T_{\rm c}$ in the BEC limit of the superfluid Fermi gas loaded on the three-dimensional optical lattice, as a function of the potential height measured in terms of the atomic recoil energy $E_r$. This figure shows the case of half-filling. $\varepsilon_{\rm F}$ is the Fermi energy. 
\label{fig9}
}
\end{figure}
%%%%%%%%%%%%%%%%%%%%%%%%%%%%%%%%%%%%%%%%%%%%%%%%%%%%%%%%%%%%%%%%%%%%%%%%%%%%%%%

\section{summary}

In this paper, we have investigated tunneling properties of a bound pair of Fermi atoms in an optical lattice. Including a realistic one-dimensional cosine-shape periodic potential, we have calculated the molecular wavefunction, binding energy, excitation spectrum, and band mass. We have also discussed validity of the Hubbard model for superfluid Fermi gases in optical lattices.
\par
In the strong coupling regime of the Hubbard model, the molecular tunneling through the lattice potential is accompanied by virtual dissociation into two atoms. This tunneling phenomenon does not actually occur in the strong coupling regime of real optical lattice system, where the bound pair moves in the lattice potential without dissociation. However, in the intermediate coupling regime, spatial structure of the molecular wavefunction indicates that the molecule dissociates into two atoms when the center of mass position is located at the top of the lattice potential. Our results show that the tunneling mechanism accompanied by virtual dissociation is realized in the optical lattice system, not in the strong coupling regime, but in the intermediate coupling regime.
\par
We have also examined the superfluid phase transition temperature $T_{\rm c}$ in the strong coupling BEC limit, where the simple Fermi Hubbard model is no longer valid. Including the correct molecular tunneling process in this limit, we showed that $T_{\rm c}$ is finite but is still low compared with the maximum $T_{\rm c}$ obtained in the intermediate coupling regime of the Hubbard model. 
\par
So far, the BCS-BEC crossover in a lattice Fermi gas has been mainly examined based on the Hubbard model consisting of the atomic hopping and on-site pairing interaction. Since this model gives vanishing $T_{\rm c}$ in the strong coupling limit, the observation of a finite and constant $T_{\rm c}$ in the BEC regime of a lattice Fermi gas would be an evidence of the pair tunneling without dissociation. Although $T_{\rm c}$ in this regime is expected to be low, the observation of the finite $T_{\rm c}$ in the BEC regime is an important challenge to clarify the validity of the Hubbard model in considering the BCS-BEC crossover regime of lattice Fermi gases. We also note that the molecular tunneling with virtual dissociation enhances the molecular mass. This naturally leads to the suppression of $T_{\rm c}$ in the intermediate coupling regime. Thus, the observation of the decrease of $T_{\rm c}$ in the intermediate coupling regime would be an indirect evidence of the virtual dissociation of the bound pair during the tunneling through the lattice potential. Since the superfluid Fermi gas in an optical lattice is an important many-body system in both cold atom physics and condensed matter physics, we expect that our results would be useful for the study of basic properties of this interesting system. 
\par
%%%%%%%%%%%%%%%%%%%%%%%%%%%%%%%%%%%%%%%%%%%%%%%%%%%%%%%%%%%%%%%%%%%%%%%%%%%%%%%
\acknowledgments
This work was supported by a Grant-in-Aid for Scientific research from MEXT, and CTC program of Japan.

%%%%%%%%%%%%%%%%%%%%%%%%%%%%%%%%%%%%%%%%%%%%%%%%%%%%%%%%%%%%%%%%%%%%%%%%%%%%%%%

\appendix
\section{Alternative derivation of Eq. (\ref{eq.5.4})}
\par
In the superfluid phase, we take $b_i=\Phi+\delta b_i$ in Eq. (\ref{eq.5.1}), where $\Phi$ is the BEC order parameter. In the mean field approximation, ignoring the fluctuation term having the form $\delta b_i^\dagger \delta b_j$, we find that Eq. (\ref{eq.5.1}) reduces to the sum of the on-site Hamiltonian as $H_M=\sum_i H_M(i)$, where
\begin{equation}
H_M(i)=zt_M\Phi^2-zt_M\Phi(b_i+b_i^\dagger)
-\mu_M n_i^M-{U_M \over 2}n_i^M(n_i^M-1).
\label{eq.a.1}
\end{equation}
Here, we have added the chemical potential term $-\mu_M n_i^M$ to Eq. (\ref{eq.a.1}). $z=6$ is the number of nearest-neighbor sites in the cubic lattice, and $\Phi$ is taken to be real. When $U_M\to+\infty$, one may only consider the vacuum state $|0\rangle$ and the single occupied state $|1\rangle\equiv b_i^\dagger|0\rangle$. Diagonalizing the on-site Hamiltonian $H_M(i)$, we obtain the eigenenergies as
\begin{equation}
E_\pm=zt_M\Phi^2-{1 \over 2}
\Bigl[
\mu_M\pm\sqrt{\mu_M^2+4(zt_M\Phi)^2}
\Bigr].
\label{eq.a.2}
\end{equation}
The free energy per lattice site is given by
\begin{equation}
F=zt_M\Phi^2-{1 \over 2}\mu_M-T\ln 
\Bigl[
2\cosh{\beta \over 2}\sqrt{\mu_M^2+4(zt_M\Phi)^2}
\Bigr].
\label{eq.a.3}
\end{equation}
The superfluid order parameter $\Phi$ is determined so as to minimize the free energy in Eq. (\ref{eq.a.4}), which gives
\begin{equation}
{\sqrt{\mu_M^2+4(zt_M\Phi)^2} \over zt_M}=\tanh{\beta \over 2}\sqrt{\mu_M^2+4(zt_M\Phi)^2}.
\label{eq.a.4}
\end{equation}
The equation for $T_{\rm c}$ is obtained by setting $\Phi=0$ in Eq. (\ref{eq.a.4}). When $n_M=1/2$, the equation for the number density of bosons is given by
\begin{equation}
{1 \over 2}={e^{\beta\mu_M} \over 1+e^{\beta\mu_M}},
\label{eq.a.5}
\end{equation}
which gives $\mu_M=0$. Substituting this result into Eq. (\ref{eq.a.4}) with $\Phi\to 0$, we obtain Eq. (\ref{eq.5.4}). 

%%%%%%%%%%%%%%%%%%%%%%%%%%%%%%%%%%%%%%%%%%%%%%%%%%%%%%%%%%%%%%%%%%%%%%%%%%%%%%%
% References


\begin{thebibliography}{99}
\bibitem{notez} In obtaining Eq. (\ref{eq.4.1}), we have taken $\Lambda_c\to\infty$. Equation (\ref{eq.4.1}) is consistent with Eq. (\ref{eq.3.4}) when $E_{\rm bind}=1/m^*a_s^2$.
\end{thebibliography}
\end{document}